\documentclass[12pt]{article}
\usepackage{graphicx,amsmath,amssymb,epsfig}

\newlength{\dinwidth}
\newlength{\dinmargin}
\def\lapproxeq{\lower .7ex\hbox{$\;\stackrel{\textstyle
<}{\sim}\;$}}
\def\gapproxeq{\lower .7ex\hbox{$\;\stackrel{\textstyle
>}{\sim}\;$}}
\def\gtrsim{\lower .7ex\hbox{$\;\stackrel{\textstyle
>}{\sim}\;$}}
\def\lesim{\lower .7ex\hbox{$\;\stackrel{\textstyle
<}{\sim}\;$}}

\def\be{\begin{equation}}
\def\ee{\end{equation}}
\def\bea{\begin{eqnarray}}
\def\eea{\end{eqnarray}}

\begin{document}
\begin{flushright}
IPPP/07/21 \\
DCPT/07/42 \\
14th May 2007 \\

\end{flushright}

\vspace*{0.5cm}

\begin{center}
{\Large \bf New Physics with Tagged Forward Protons at the LHC\footnote{
Presented 
by V.A.~Khoze
at the 21st Workshop Les Rencontres de Physique de la Vall\'{e}e d'Aoste, La Thuile, 
Aosta Valley, 4-10 March, 2007.}} \\   

\vspace*{1cm}
\textsc{V.A.~Khoze$^{a,b}$, A.D. Martin$^a$ and M.G. Ryskin$^{a,b}$} \\

\vspace*{0.5cm}
$^a$ Department of Physics and Institute for
Particle Physics Phenomenology, \\
University of Durham, DH1 3LE, UK \\
$^b$ Petersburg Nuclear Physics Institute, Gatchina,
St.~Petersburg, 188300, Russia \\

\end{center}

\vspace*{0.5cm}

\begin{abstract}
The addition of forward proton detectors to LHC experiments will significantly
enlarge the potential for studying New Physics.
A topical example is
Higgs production by the central exclusive diffractive process, $pp \to p+H+p$.
We discuss the exclusive production of Higgs bosons in both the SM and MSSM.  
Special attention is paid to the backgrounds to the $H \to b\bar b$ signal.
\end{abstract}

\newpage
\section{Introduction}
\begin{figure} [b]
\begin{center}
\includegraphics[height=3cm]{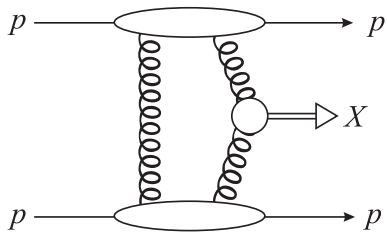}
\caption{The basic mechanism for the exclusive process $pp \to p+X+p$. The system $X$ is produced by the fusion of two active gluons, with a screening gluon exchanged to neutralize the colour.}
\label{pxp} 
\end{center}
\end{figure}

The use of diffractive processes to study the Standard Model (SM) and New Physics
at the LHC has only been fully appreciated within the last few years; see, for
example
 \cite{KMR,INC,DKMOR,cox1}, or the recent reviews \cite{jeff,KMRrev1,KMRrev2}, and references therein.
By detecting protons that have lost only about 1-3$\%$  of their longitudinal
momentum \cite{ar,fp420}, a rich QCD, electroweak, Higgs and BSM programme becomes accessible experimentally, with
the potential to study phenomena which are unique to the LHC, and difficult even at a future
linear collider.
Particularly interesting are the so-called central exclusive production (CEP)
 processes which provide an extremely favourable environment to search for, and identify 
the nature of, new particles at the LHC. The first that comes to mind are the Higgs bosons, but
there is also a potentially rich, more exotic, physics menu 
including (light) gluino and squark production, searches for extra dimensions,
 gluinonia, radions, and indeed any new
 object which has $0^{++}$  (or $2^{++}$) quantum 
numbers and couples strongly to gluons, see for instance \cite{INC,KPR,jeffg}. 
By ``central exclusive'' we mean a process of the type $pp\rightarrow p + X + p$, where 
the + signs denote the absence of hadronic activity 
(that is, the presence of rapidity gaps) between the outgoing protons and the 
decay products of the centrally produced system $X$. The basic mechanism driving the
process is shown in Fig.~\ref{pxp}.

There are several reasons why CEP is especially attractive for searches for
new heavy objects.
{\it First}, if the outgoing protons remain
 intact and scatter through small angles then, to a very good approximation,
the primary active di-gluon system obeys a $J_z=0$, C-even, P-even, selection rule
\cite {KMRmm}. Here $J_z$ is the projection of the total angular momentum
along the proton beam axis. This selection rule readily permits a clean determination 
of the quantum numbers of the observed new
(for example, Higgs-like) resonance, when the dominant production is a scalar state.
{\it Secondly}, because the process is exclusive, the energy
loss of the outgoing protons is directly related to the mass of the
central system, allowing a potentially excellent mass resolution,
irrespective of the decay mode of the
centrally produced system.
 {\it Thirdly}, in many topical cases, in particular,
for  Higgs boson production,
 a signal-to-background ratio of order 1
(or even better) is achievable \cite{DKMOR,jeffg}, \cite{KKMRext}-\cite{clp}.
In particular, due to $J_z=0$ selection, 
leading-order QCD  $b\bar b$
 production is suppressed by a 
factor $(m_b/E_T)^2$, where $E_T$ is the transverse energy of the $b,\bar b$ jets.
 Therefore, for a low mass Higgs, $M_H \lapproxeq 150$ GeV,
 there is a possibility to observe the 
main $b\bar b$ decay mode \cite{INC,DKMOR,KMRrev1},
and to directly measure the $H \rightarrow b\bar b$ Yukawa coupling constant. 
The signal-to-background ratio may become significantly larger for a 
 Higgs boson in certain regions of the MSSM parameter space 
\cite{KKMRext,HKRSTW}.

It is worth mentioning that,
by tagging both of the outgoing protons, 
the LHC is effectively turned into a gluon-gluon collider.
This will open up a rich, `high-rate' QCD physics menu (especially
concerning diffractive phenomena), 
which will allow the study of the skewed, unintegrated
gluon densities, as well
as the details of rapidity gap survival; see, for example, \cite{INC,KMRrev2,maor}. 
Note that CEP provides a source of practically
pure gluon jets; that is we effectively have a `gluon factory' \cite{KMRmm}. 
This provides an ideal laboratory in which
to study the detailed properties of gluon jets, especially in comparison with
quark jets.
The forward-proton-tagging  approach also offers a unique programme of
high-energy photon-interaction physics at the LHC; see, for example, \cite{kp,KMRphot}.

\section{Central Exclusive Higgs production}
The `benchmark' CEP  new physics
process is Higgs production. Studies of the Higgs sector are at the heart
of the recent proposal \cite{fp420} to complement the LHC central
detectors with proton taggers placed at 420 m either side of the interaction point.

 Our current understanding is, that
if a SM-like Higgs boson exists in Nature, it will be detected at the
LHC. However, various extended models predict a large diversity of Higgs-like
bosons with different masses, couplings and  CP-parities. 
The best studied extension of the SM up to now is the Minimal 
Supersymmetric Standard Model (MSSM) \cite{CH}, in which there are
three neutral Higgs bosons, the scalars $h$ and $H$, and the pseudoscalar $A$.

The forward proton tagging mode
is especially advantageous for the study of the MSSM sector \cite{KKMRext,HKRSTW}. 
Note that when  using the "standard" non-diffractive
production mechanisms, there is usually
an important region of MSSM parameter region, where the LHC can
detect only the 
 Higgs boson with SM-like properties.  
To check that a discovered state is indeed a scalar Higgs boson, and to
distinguish between the Higgs boson(s) of the SM or the MSSM and those from
of extended Higgs theories will be highly non-trivial task. 
Without forward proton tagging, it would
require interplay with observations at the Next Linear Collider.
Moreover, within the MSSM, the weak-boson-fusion
channel becomes of no practical use for the production of the heavier
scalar $H$ or the pseudoscalar $A$ boson.
On the other hand, in the forward proton mode
the pseudoscalar $A$ is practically filtered out, and the
detection of the $H$ boson 
 should be achievable \cite{KKMRext,HKRSTW}. 
In addition, in some MSSM  scenarios, CEP provides
an excellent opportunity for probing 
the CP-structure of the  Higgs sector,
either by measuring directly the azimuthal asymmetry 
of the outgoing tagged protons 
\cite{KMRCP} or by studying the correlations between
the decay products \cite{je}.

\begin{figure}
\begin{center}
\includegraphics[height=10cm]{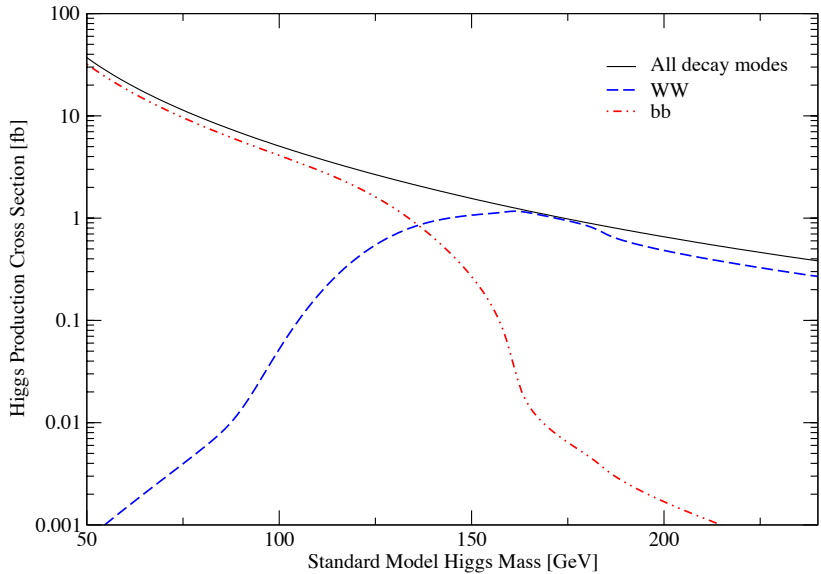}
\caption{
The cross section times branching ratio for CEP of the SM Higgs \cite{cox2}.}
\label{exfigtot} 
\end{center}
\end{figure}

In Fig.~$\ref{exfigtot}$ we show, for reference purposes, the total CEP cross section for the SM Higgs boson times
branching ratio for the
$WW$ and $b\bar b$ channels, as a function  of the Higgs mass.
We see that the expected total cross section for the CEP of 
a SM Higgs, with mass 120 GeV, is 
3~fb, falling to just less than 1~fb for a mass of 200 GeV; see \cite{KMR}.

With a good understanding of the detectors and  
 favourable experimental conditions, the
rate for the SM Higgs of mass 120 GeV for the integrated
 LHC luminosity of 
${\cal L} = 60~{\rm fb}^{-1}$ would be quite sizeable
(around 100 events). However,
with the presently envisaged LHC detectors, there are various experimental
problems. First of all, trigger signals from protons detected at 420 m
cannot reach the central detector in time to be used in the Level 1 trigger. For this,
we have to rely on the central detector.
Other factors may also strongly reduce the current expectations
for the detected signal  rate, in particular,  the $b$-tagging efficiency, the
jet energy resolution etc. At high luminosities there is 
also a potentially dangerous problem
of backgrounds due to the overlapping events in the same bunch crossing
(the so-called ``pile-up'' events). In summary, with the current  hardware, the expectation is that there will be not more than a dozen SM Higgs signal events for an integrated
LHC luminosity of ${\cal L} = 60~{\rm fb}^{-1}$. Whether experimental ingenuity will increase this number remains to be seen.
Indeed, it is quite possible that ``clever'' hardware and the use of optimized cuts will increase the rate. 
For example, the number of  $h \to WW^*$ events would double  if the
trigger thresholds on single leptons could be reduced \cite{cox2}.
Further improvement of the $b$-tagging efficiency and
of the jet energy resolution would be particularly welcome.
Note that the forward-proton mode offers the possibility
to study  the combined event rate using the so-called 'trigger cocktail'. 

\begin{figure}
\begin{center}
\includegraphics[height=8cm]{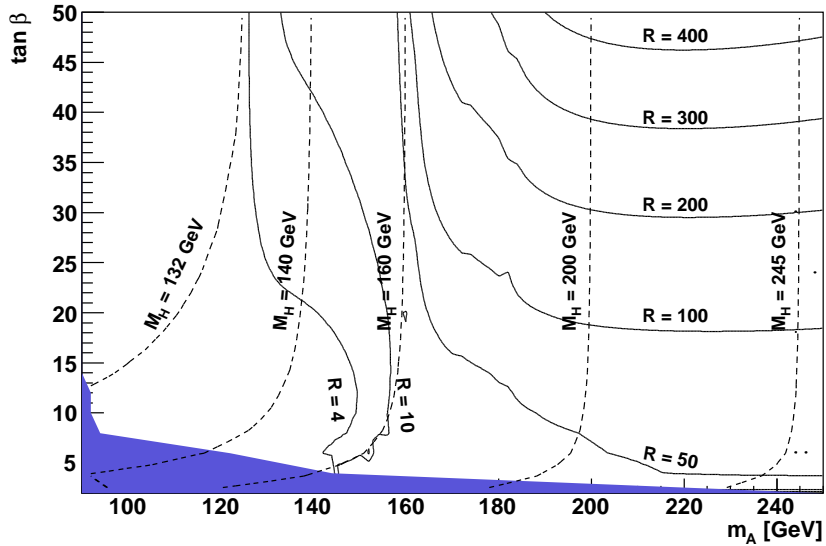}
\caption{Contours for the ratio $R$ of the $H \to b \bar b$ signal events in the MSSM over those in the SM in 
CED process in the $M_A$--$\tan\beta$ plane. The ratio is 
shown for the $M_h^{\rm max}$ benchmark scenario (with $\mu = +200$ GeV).
The values of the mass of the lighter CP-even Higgs boson, $M_h$, are
indicated by dashed contour lines. The dark shaded 
region is excluded by the LEP Higgs searches.}
\label{ratio}
\end{center}
\end{figure}

As we already mentioned, in the MSSM, the CEP cross sections can be an order-of-magnitude or more
higher \cite{HKRSTW}. This is illustrated in Fig.~\ref{ratio}, which shows the 
contours for the ratio $R$ of signal events in the MSSM over those in the SM in 
the  CEP of $H \to b \bar b$ in the $M_A$--$\tan\beta$ plane, see \cite{HKRSTW}.



As discussed above, the exclusive Higgs signal 
is particularly clean, and the signal-to-background
ratio is  quite favourable, at least, at an instantaneous
luminosity ${L}\sim 2 \times 10^{33} \, {\rm cm}^{-2} \, {\rm s}^{-1}$, 
when the effect of pile-up 
 can be kept under control, see 
\cite{cms-totem, HKRSTW}. However, without improving
the LHC hardware, the expected event rate in the 
SM case is quite limited, and so it is important to test various
ingredients of the
adopted theoretical scheme \cite{KMR, KMRmm,INC}
 by  studying the related processes at the 
existing experimental facilities, HERA and the Tevatron.
Various such tests have been performed so far, see for
example, \cite{KMRrev1,KKMR,AKR} and references therein. Quite recently the
predictions for the non-perturbative so-called survival
factor have been confronted with HERA data on the leading neutron
spectra \cite{leadneutr}. 

The  straightforward checks
come from the study of processes which are mediated  
by the same mechanism as CEP of the Higgs boson, but
with rates which are sufficiently high, so that they may be observed at the Tevatron
(as well as at the LHC).  The most obvious examples are those in which the Higgs 
is replaced by either a dijet system, or a $\chi_c$ meson, or a $\gamma \gamma$ pair.
The reported preliminary CDF data on these CEP processes
(see for example,\cite{mg,dino,koji}) show a reasonable agreement
with the theoretical expectations by Durham group, see also \cite{royon}.

\begin{figure} [h]
\begin{center}
\centerline{\epsfxsize=0.6\textwidth\epsfbox{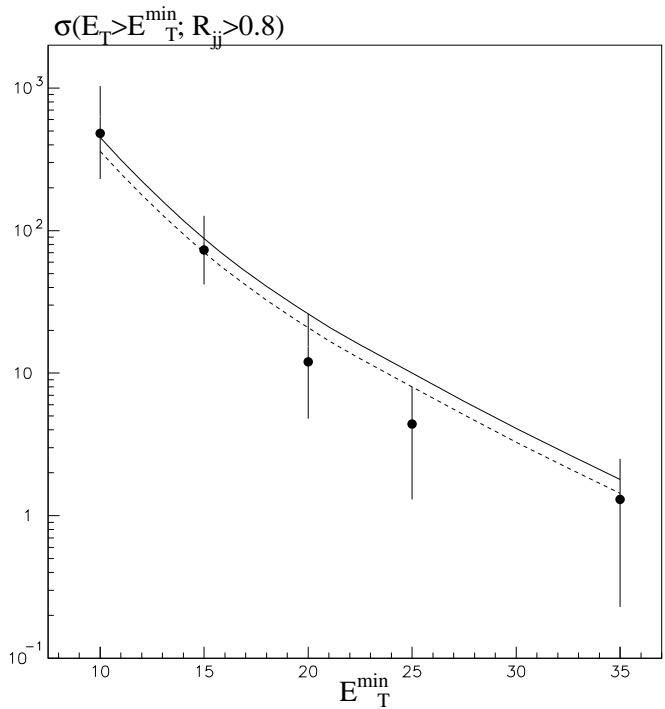}}
\caption{The cross section  for `exclusive' dijet production
 at the Tevatron as a function $E_T^{\rm min}$ as measured by CDF \cite{dino}. 
  These preliminary CDF data correspond to the cross section integrated over
the domain $R_{jj} \equiv M_{\rm dijet}/M_{\rm PP} > 0.8$ and $E_T > E_T^{\rm min}$. A jet cone of $R<0.7$ is used.
The curves are the pure exclusive cross section
calculated \cite{INC} using the CDF event selection. 
The solid curve is obtained by rescaling the parton (gluon)  transverse momentum
$p_T$ to the measured jet
transverse energy $E_T$ by  $E_T=0.8 p_T$. The dashed curve assumes
$E_T=0.75 p_T$. The rescaling procedure effectively accounts for the hadronization
and radiative 
effects, and for jet energy losses outside the selected jet cone.
This prescription
for parton jet energy loss is in agreement with the
 out-of-cone energy measurements in CDF \cite{koji1}.}
\label{fig:JJ}
\end{center}
\end{figure}

Especially
impressive are the recent CDF data \cite{dino, koji} on
 exclusive production
of a pair of high $E_T$ jets, $p\bar {p} \to p+jj+\bar {p}$.
As discussed in \cite{KMR,INC} such measurements could provide
 an effective $gg^{PP}$ `luminosity
monitor' just in the kinematical region appropriate for Higgs
production. The corresponding cross section was evaluated to
be about 10$^4$ times larger than that for the production of a SM Higgs boson.
Since the dijet CEP cross section is rather large, this  process appears
to be an ideal `standard candle'.
A comparison of the data with 
analytical predictions \cite{KMR,INC}  is given in Fig.~$\ref{fig:JJ}$. 
It shows the  $E_T^{\rm min}$ dependence  
for the dijet events with $R_{jj} \equiv M_{{\rm dijet}}/M_{\rm {PP}} > 0.8$,
where $M_{\rm {PP}}$ is the invariant energy of the incoming
Pomeron-Pomeron system.  The agreement with the theoretical expectations
\cite{KMR,INC} lends credence to the predictions for the CED
Higgs production \cite{dino}.

\section{The backgrounds to the $p+(h, H\to b\bar b)+p$ signal}
The importance of the $p+(h, H\to b\bar b)+p$ process, in particular
as a SUSY Higgs search mode, means that the physical backgrounds to this reaction
must be thoroughly addressed. 
Recall that the unique advantage of the  $b\bar b$ CEP process is   
the $J_z=0$ selection rule, which requires the
LO $gg^{PP}\to b\bar b$ background  to vanish in
the limit of massless quarks and forward going protons\footnote{This is an example of the so called Maximally Helicity 
Violating (MHV) rule, see for review \cite{MP}.}.
However, there are still four main sources of
background 
\cite{DKMOR,KMRrev1,krs2}. These are the contributions from the following subprocesses.
\begin{itemize}
\item[(i)] 
The prolific (LO)
$gg^{PP}\to gg$ subprocess can mimic $b\bar b$ production since we may
misidentify the  gluons as $b$ and $\bar{b}$ jets.

\item[(ii)]
An admixture of $|J_z|=2$ production, arising from non-forward going
protons, which contributes to the (QHC\footnote{It is convenient to consider
separately the  quark helicity-conserving (QHC) and the quark helicity-non-conserving (QHNC) amplitudes \cite{krs2}. These amplitudes do not
interfere, and their contributions can be treated independently.}) LO $gg^{PP}\to b\bar b$ background. 
 
\item[(iii)] 
Because of non-zero mass of the quark there is a contribution to the $J_z=0$ 
(QHNC) cross section of order $m_b^2/E_T^2$. This term  currently 
raises the main concern. 
The problem is that the result is strongly affected by the 
(uncomfortably large) higher-order QCD effects 
 see \cite{fkm,krs2}. 
In particular, the one-loop double logarithmic 
contribution exceeds the Born term, and the final result
becomes strongly dependent on the NNLO  effects, as well as
on the scale $\mu$ of the QCD coupling $\alpha_S$
 and on the running $b$~quark mass.  
There is no complete calculation of these higher-order 
effects for the $gg^{PP}\to b\bar b$ process, but only estimates
based on a seemingly plausible hypotheses regarding the NNLO effects \cite{krs2}.
The validity of these
estimates has an 
accuracy not better than a factor of 2-4. This contribution is the main source of the 
theoretical uncertainty in the current predictions for the non-pile-up
background. The good news is that this contribution decreases with 
increasing $E_T$ much faster than the other background terms
\cite{KKMRext,KMRCP}. 

\item[(iv)] 
Finally, there is a possibility of NLO $gg^{PP}\to b\bar b g$
background contributions, which for large angle, hard
gluon radiation do not obey the selection rules, 
see \cite{DKMOR, krs2}.
Of course, in principle, the extra gluon may be observed
experimentally and the contribution of such background events
reduced. 
However, there are important exceptions \cite{DKMOR,krs2}.
First, the extra gluon may go unobserved in the direction of
a forward proton. This background is reduced by
requiring the approximate equality $M_{\rm missing} = M_{b\bar b}$.
Calculations \cite{kmrN} show that this  background does not exceed  
5$\%$ of the SM Higgs signal, and so it may be safely neglected.
The remaining danger is large-angle hard gluon emission which is
collinear with either the $b$ or $\bar{b}$ jet, and, therefore,
unobservable. 
This background source results
in a sizeable contribution which should be included, see \cite{HKRSTW}. 
\end{itemize}


There are also other (potentially worrying) background sources,
which after a thorough investigation \cite{krs2,kmrN}, have been omitted in the final expression for the  $b \bar b g$ background in \cite{HKRSTW}. This is either because their
contributions are numerically small from the very beginning,
or because they can be reduced to an acceptable level 
by straightforward experimental cuts.
Among these, there is the NNLO QHC (``cut non-reconstructible'')
 contribution to the
exclusive process, which comes from the one-loop box diagrams.
This contribution is not mass-suppressed and is potentially important, 
especially for large $M_H$. 
However, for masses below 300 GeV, this contribution is comparatively small.

Next, a potential background source can arise from  
the collision of two soft Pomerons. This can result in
the two main categories of events:
\begin{itemize}
\item[(a)] central Higgs boson production accompanied by
two (or more) additional gluon jets,

\item[(b)] production of a high $E_T$ $b\bar b$-pair accompanied by the gluon
jets.
\end{itemize}
In these cases the Higgs boson or the $b\bar b$ pair are produced in the
collision of two gluons (from the Pomeron wave functions) via the hard
subprocesses ($gg\to H$ or $gg\to b\bar b$) similar to the usual
inelastic event.
In both processes the mass, $M_{bb}$, of the central  $b\bar b$ system
(resulting either from the Higgs decay or from the QCD background)
is not equal to the `Pomeron-Pomeron' mass $M_{PP}=M_{\rm missing}$,
measured  by the  proton detectors.
The suppression of such backgrounds
is controlled by the requirement that $|M_{\rm missing}-M_{bb}|$ should
lie within the $\Delta M_{bb}$ mass interval.
These backgrounds were carefully evaluated in \cite{kmrN}, and it was found that they are quite small. Indeed, if we use the MRW2006 DPDFs \cite{mrw}, and take $\Delta M_{bb}\simeq 24$ GeV, then the $gb\bar bg$ and $gHg$ contributions 
are each less than about  6$\%$ of the SM Higgs signal.

Finally, a potential background could result from the emissions of
additional gluons.
A particular case, caused by 
the QCD $b\bar b + {\rm gluons}$ process, was already addressed in the
item (iv) above. There may also be a contribution coming from
the  $H+ng$  production process. This contribution is suppressed by 
the requirement that the $t$-channel two-gluon exchange
across the gap region should be colourless. Thus, there is no single gluon radiation, and the non-zero contribution
starts from $n=2$. Next, we have to impose the mass matching condition
discussed in the item (iv) above.
Numerically, this background appears to be small 
(about 15$\%$ of the SM Higgs signal \cite{kmrN}) 
and, again, it can be neglected.
It should be noted
that the effect of gluon emission off the screening gluon (see Fig.~\ref{pxp})
is also numerically small.

In summary, the main background contributions come from
exclusive dijet production as listed in the items (i)-(iv) above.
Within the accuracy of the existing calculations \cite{KMRmm,DKMOR,krs2},
the overall background to the $0^+$ Higgs signal in the 
$b \bar b$ mode can be approximated by the following
formula, see \cite{HKRSTW}
\be
\frac{{\rm d}\sigma^B}{{\rm d} M} \approx 0.5 \, {\rm fb/GeV} \left[
0.92\left(\frac{120}{M}\right)^6 +
\frac{1}{2} \left(\frac{120}{M}\right)^8
                                  \right],
\label{eq:backbb}
\ee
where the first term in the square brackets corresponds to
the processes listed in items (i), (ii) and (iv), while the last term 
comes from the mass-suppressed term described in item (iii).
We emphasize that this approximate expression may be used
only for the purposes of making quick estimates  
of the background, since
no detector simulation has been performed.  We expect that such a 
simulation, together with the optimization procedure, will further
reduce the effect of background.

\section{Detecting the exclusive Higgs $\to WW$ signal}
Although the $H \to b \bar b$ signal has special advantages, we have discussed  
problems which arise, in the SM case, to 
render it challenging from an experimental perspective.
In \cite{cox2,krs1}, attention was turned to
 the $WW$ decay mode. 
Triggering on this channel is
not a problem, since the final state is rich in high-$p_{\rm T}$
leptons. Efficiencies of
about 20$\%$ can be achieved if the standard
leptonic and di-leptonic trigger thresholds are applied. 
The advantages of forward proton tagging are, however, 
still explicit. Even for the
gold-plated double leptonic decay channel,
 the mass resolution will be very good, and, of course, the 
observation of the Higgs with the tagged protons immediately 
establishes its quantum numbers.

It was
demonstrated in \cite{cox2,krs1} that there would be
a detectable signal with a small and controllable background
for the CED production of a SM-like Higgs boson in the mass interval
between 140 GeV and 200 GeV.
Unfortunately, with the standard lepton triggers and
experimental acceptances and selections~\cite{cox2}, currently we can expect only a handful of $WW^*$ events
from a 120 GeV SM Higgs for ${\cal L}=60~{\rm fb}^{-1}$.
The rate of detected events could rise after further 
modifications of hardware. For example, the reduction of the  Level~1 leptonic trigger thresholds
would allow the statistics to double.
As shown in \cite{HKRSTW}, 
the situation would improve in favourable
regions of the MSSM parameter space, but here, unlike the $b \bar b$ mode, the expected rise, as compared to SM, is not dramatic, no more
than a factor of 4-5.
In order to fully exploit all the advantages of the 
$WW$ channel more dedicated experimental studies are needed.

\section{Conclusion}

The installation of 
 proton-tagging detectors in the distant forward regions around the 
ATLAS and/or CMS central detectors would add unique capabilities 
to the existing LHC experimental programme. The calculation of the rates of CEP
processes show that 
 there is a real chance that new 
 heavy particle production could
 be observed in this mode. For a 
Higgs boson this would amount to a direct determination of its quantum numbers. 
For certain MSSM scenarios, the tagged-proton channel may even be 
the Higgs discovery channel. Moreover, with sufficient luminosity, proton tagging 
may provide direct evidence of CP-violation within the Higgs sector. There is also a rich
QCD, electroweak, and 
more exotic physics, menu. This includes searches for extra dimensions, light
gluino and 
squark production, gluinonia, radions, and, indeed, any object which has $0^{++}$ or $2^{++}$ quantum 
numbers and which couples strongly to gluons \cite{INC}.
 
Here we focused on the unique advantages of CEP
Higgs production. 
The events are clean, but the predicted yield   for the SM Higgs
 for an integrated luminosity of ${\cal L}=60~{\rm fb}^{-1}$ is comparatively low, after experimental cuts and acceptances. Further efforts
to optimize the event selection and cut procedure are very desirable. 
 The signal-to-background ratio in  the $b \bar b$ mode is about 1,
depending crucially on the accuracy with which $M_{\rm missing}$ can be measured.
In the MSSM there are certain regions 
of parameter space which can be especially `proton tagging friendly'
\cite{KKMRext,HKRSTW}. Here the signal-to-background ratios in the $b \bar b$ channel
can exceed the SM by up to two orders of magnitude.
Moreover, the observation of the decay of Higgs to $b \bar b$ would  allow a direct
determination of
the $b$~Yukawa coupling.

From the experimental perspective, the simplest exclusive channel in which to observe a 
SM Higgs boson with mass between
140 GeV and 200 GeV
is the $WW$ decay mode. According to studies in \cite {cox2},
 there will be a detectable signal at ${\cal L}=60~{\rm fb}^{-1}$, and the
 non-pile-up  backgrounds
are small and controllable. 
However, contrary to the $b \bar b$ case, 
 no dramatic rise in  the rate is expected within the MSSM.

 Potentially, the pile-up events could endanger the prospects of CEP
studies at high luminosities. Currently the situation is far
from being hopeless, but further detailed studies are needed.
The pile-up  is currently under very intensive scrutiny by both, ATLAS
and CMS; for a detailed discussion, see \cite{cms-totem}, (see also \cite{clp,HKRSTW}).

\section{Acknowledgements}
We thank 
Brian Cox, Albert De Roeck, Sven Heinemeyer,
 Risto Orava, James Stirling, Andy Pilkington, Marek Tasevsky,
Koji Terashi 
and Georg Weiglein
for useful discussions. 
VAK is very grateful to Giorgio Bellettini,
 Giorgio Chiarelli and Mario Greco for the
 kind invitation and warm hospitality at  La Thuile.
This work was supported by
INTAS grant 05-103-7515, by grant RFBR 07-02-00023,
by Federal Program of Russian Ministry of Industry,
Science and Technology RSGSS-5788.2006.02.

\end{document}